# Superlubric Schottky Generator in Microscale with High Current Density and Ultralong Life

Xuanyu Huang[1], Xiaojian Xiang[1], Deli Peng[1], Fuwei Yang[1], Haiyang Jiang[1], Zhanghui Wu[1], Zhiping Xu[1], Quanshui Zheng[1,2*]

[1] Department of Engineering Mechanics and Center for Nano and Micro Mechanics, Tsinghua University, Beijing 100084, China

[2] State Key Laboratory of Tribology and Applied Mechanics Lab, Tsinghua University, Beijing 100084, China,

*E-mail: zhengqs@tsinghua.edu.cn

**Miniaturized or even microscale generators that could effectively and persistently converse weak and random mechanical energy from environments into electricity promise huge applications in the internet of things, sensor networks, big data, personal health systems, artificial intelligence, etc. However, such generators haven't appeared yet because either the current density, or persistence, or both of all reported attempts were too low to real applications.** Here, we demonstrate a superlubric Schottky generator (SLSG) in microscale such that the sliding contact between a microsized graphite flake and an n-type silicon is in a structural superlubric state, namely a ultralow friction and wearless state. This SLSG generates a stable electrical current at a high density (~119 $Am^{-2}$) for at least 5,000 cycles. Since no current decay and wear were observed during the entire experiment, we believe that the real persistence of the SLSG should be enduring or substantively unlimited. In addition, the observed results exclude the mechanism of friction excitation in our Schottky generator, and provide the first experimental support of the conjectured mechanism of depletion layer establishment and destruction (DLED). Furthermore, we demonstrate a physical process of the DLED mechanism by the use of a quasi-static semiconductor finite

**element simulation. Our work may guide and accelerate future SLSGs into real applications.**

With the rapid development of nanotechnology and microfabrication technology, ceaselessly miniaturized sensors and devices are emerging in vast numbers of applications in the internet of things, sensor networks, big data, personal health systems, artificial intelligence, etc. [1-5]. Until now, these sensors and devices have been mostly powered by line cords, or by batteries and external chargers, and thus have been limited in applications [6,7] particularly in needs for independent, sustainable, maintain-free operations of implantable biosensors, remote and mobile environmental sensors, nano/micro-scale robots, portable/wearable person electronics, etc [7-10]. Therefore, the above devices require new power supplies with general characteristics that are small, wireless, portable, and sustainable.

As a promising solution to the above challenge, nanogenerators were proposed firstly in 2006 [1] to converse weak and random mechanical energy from the environment and biological systems into electricity. Since then, many kinds of nanogenerators have been proposed, such as piezoelectric nanogenerators (PENGs) [1,11], triboelectric nanogenerators (TENGs) [12], and electret-based microgenerators (EBMGs) [13]. More recently, a new kind of nanogenerators, tribological Schottky generators (TSGs) that are based on relatively sliding Schottky junctions and the mechanism of friction excitation, has been proposed [14-17]. Compared with previously proposed nanogenerators, TSGs have the advantages that can generate DC currents with much higher current densities and meanwhile have simpler structures, in addition to flexible material choices [3,14,16]. However, to the best of our knowledge all real setups of nanogenerators reported in the literature don't simultaneously have a sufficiently high current density and a sufficiently long life for real applications [3,14]. A fundamental, maybe the biggest, challenge is the contradiction between output current density and life due to the existence of friction and wear. For instance, the reported highest current densities were in the range of $10^4 \sim 10^7$ $Am^{-2}$, generated by TSGs through rude frictions (with respect to normal pressures of $1 \sim 10$ GPa) of nanotips on semiconductor substrate [3,16,18]. The corresponding lifes were extremely short (1~5 circles) [3,18]. On the other hand, the reported longest lifes of TSGs were in the ranges of 3,600~10,000 circles [14,19], that correspond to slight frictions (with respect to normal pressures of 2~5 kPa) and generate too low current densities of 0.0055~0.13 $Am^{-2}$ to most applications [14,19-21].

Structural superlubricity (SSL) is a state of ultrallow friction and wearless between two solid surfaces [22]. Since the first realization of microscale SSL in the atmospheric environment in 2012 [23], in addition to the realizations of high speed SSL (25 m/s ~ 293 m/s) [24,25], SSL has aroused interest in practical application, bringing a dawn for the revolutionary solution of friction and wear probems [22]. As an application of SSL, several types of superlubric nanogenerators (SLNGs) based on capacitors or electrets were theoretically proposed recently by the authors [26]. The results show that SLNGs promise to acheve the maximum electric current density of ~150 $Am^{-2}$ at a sliding speed of 1 mm/s, that is three orders of magnitude higher than the maximum density of ~0.1 $Am^{-2}$ of all reported TENGs [27] and PENGs [28], and have nearly 100% conversion efficiency and superb long lifetime. In this Letter, we demonstrate a superlubric Schottky generator (SLSG) as the first physical prototype of superlubric nanogenerators, which base on a new mechanism mostly like the depletion layer establishment and destruction (DLED) mechanism instead of friction excitation mechanism, that can generate a stable and high current density of ~119 $Am^{-2}$ for at least 5,000 cycles.

We take three steps to achieve the SLSG prototype. At the first step, we select a graphite flake that was cleaved by a shear from a square graphite mesa of the dimensions 4 μm × 4μm × 2.6μm and has the self-retracting motion (SRM) property [29] (the Methods section and Supplementary Section 2 gives further details).

The graphite mesa was made of a HOPG (highly ordered pyrolytic graphite) with an Au film of thickness 100 nm (the fabrication details are shown in Supplementary Section 1 and Method section). It is known that the cleaved surface (or the bottom surface) of a SRM flake is a single crystalline graphene sheet and is superlubric [30].

At the second step, as illustrated in Fig. 1a, we transfer the above flake onto an atomic smooth surface of an n-type silicon with the doping concentration $N_D = 10^{15} cm^{-3}$ that is coated with an Al electrode on the back of the film (the Method section gives the fabrication process). Because of the different work functions of the graphite flake and the n-type Silicon (n-Si) film, electrons will automatically transfer between the graphite and n-Si contacted surfaces to form a depletion layer, and therefore form a Schottky barrier when it reaches equilibrium (the Supplementary Fig. S3 gives its I-V characteristic curve).

At the third step, we apply a normal force, $N$, on the top of the graphite-Au flake by a conductive AFM tip (coated with Au), which will generate a friction force, between the tip and the Au film. Then, we induce a relative sliding between the superlubric graphite surface and the smooth n-Si surface through a lateral motion of the piezoelectric displacement platform, as illustrated in Fig. 1b by the optical image of the experimental setup. A current measurement circuit is connected to the conductive AFM tip and the Al electrode coated on the bottom of the n-Si. When relatively sliding the graphite flakes against the n-Si, we measure the current in the external circuits, and meanwhile measure the friction force by the AFM's lateral force mode.

The measured current is plotted in Fig. 1c for the first 2000 cycles as we relatively slide the graphite flake with displacement amplitude 2μm and speed 4μm/s under a normal force of $N = 22.3$ μN on the graphite flake (the Method section and Supplementary Section 3 gives further details, and the Supplementary video 1 gives the experimental demonstration). The results show that the current during each sliding cycle is nearly constant, and the current gradually increases with the number of sliding cycles (the noise current measurement in Supplementary Fig. S5, and the decay of the current after the graphite flake layered in Supplementary Fig. S6 confirms that the current is caused by the sliding of the graphite/n-Si interface). After that, we did a longer (around 5000 cycles) sliding test with the same normal force $N = 22.3$ μN, but stepwise-increasing the sliding speed from 4μm/s to 24μm/s. As plotted in Fig. 1d, the measured results show that the current (blue line), $I$, increases with the number of sliding cycles under different sliding speeds, the maximum current can reach around 1.9 nA, and the maximum current density is 119 A/m² for the contact area of 4μm × 4μm, that is similar high as the maximum possible current density (~150 Am⁻²) at sliding speed of 1 mm/s of superlubric capacity nanogenerators allowed before electric breaking [26]. In addition, the measured friction force (red line) has a decline process [31] in the first around 64 cycles, and has basically stabilized in the subsequent thousands of slides under different sliding speed (only slowly increasing with the sliding speed). Later on we shall show that the mechanism of generating electicity in the above-demonstrated SLSG is very impossible to be the friction excitation mechanism, but is most-likely the depletion layer establishment and destruction (DLED) mechanism [14,15].

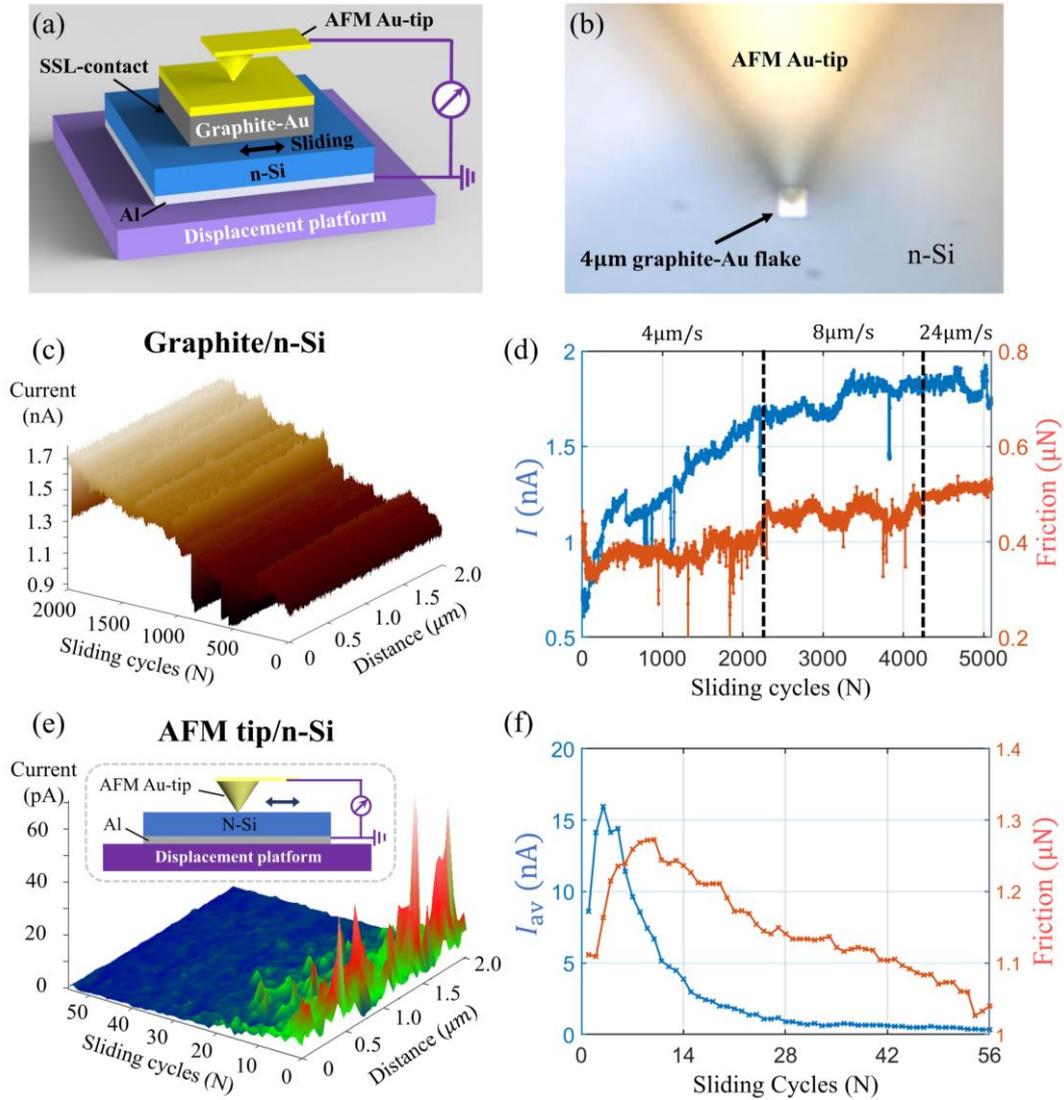

**Fig. 1 | The output current and friction measurement of SLSG and the comparison with TSG. a**, The structure of graphite/n-Si SLSG. **b**, The optical microscope observation of graphite/n-Si SLSG. **c**, The current maps in first 2000 cycles within grapite/n-Si SLSG. **d**, The relationship between friction and current $I$ with sliding cycles of graphite/n-Si SLSG in difference speed. **e**, The current maps in first 56 cycles within AFM tip/n-Si TSG, and the illustration is the experiment setup of AFM tip/n-Si TSG. **f**, The relationship between friction and average current $I_{av}$ with sliding cycles of AFM tip/n-Si TSG.

As a comparison, an experiment of sliding an AFM Au-tip of radius of curvature 35 nm with the normal force 4.49 μN pressed on the same n-Si film is performed (the Method section and Supplemetary gives the futher details). The measured current versus sliding circles is plotted in Fig. 1e, with the insert indicating the experimental setup. The results show that the current of the AFM tip/n-Si tribological Schottky generator (TSG) is pulse, unlike the stable output of the graphite/n-Si SLSG shown in Fig. 1c. The average current, $I_{av}$, per circle and the friction force versue circles are depicted in Fig. 1f. Although the $I_{av}$ the tip/n-Si TSG can reach its momentary maximum value of around 15pA, it decays very quickly to ~1 pA at about 30ᵗʰ circle, which is three orders lower than that of SLSG at 5000ᵗʰ circle. On the other hand, the momentary maximum average current density can reach at ~$4.0 \times 10^4$ Am⁻² if we estimate the contact area of the tip as $A = $ ~340 nm² according to the Derjaguin–Muller–Toporov (DMT) contact model [32] (the Method section gives the futher details), which is two orders of magnitude larger than that of the SLSG (~119 Am⁻²).

The above results are consistent with the reported very high current densities ($10^4$~$10^7$ Am⁻²) and extralow lifes (1~5 circles) in the literature [3,18] for all TSGs made of nanotip and semiconductor contacts, which reflect the nature of friction excitation. Attempts for high circle number tests of TSGs were made in terms of macroscale solid-solid contacts, showing stable current densities (0.0055~0.13 Am$^{-2}$) for large circles upto 3,600 ~ 10,000 [14,19]. However, the large circles are result of very low applied normal pressures (2~5 kPa), that correspond to low wear and pay the cost of low friction excitation [14,19-21]. In all reported TSGs [3,14,16,18-21,33-35], serious of obvious wears were observea, and the mechanisim was mostly explained as the friction excitation [15-17]. In this mechanism, during a sliding electrons or holes are generated due to the energy released by sliding friction and bonding interaction at the contacted surfaces [35], which changes the Fermi level of the semiconductor, and electrons or holes would be expelled out of the Shottky junction due to the built-in electric field to form DC current. As a consequence, the generated current density should increase with normal pressures or friction forces, as well as sliding speeds. These properties were experimentally confirmed [20], that support the friction excitation mechanism.

In contrast, in the SLSG reported above the contact between the graphite flakes and n-Si in the SLSG is in the state of structural superlubricity (SSL). To verify this, we conducted a series of tribological tests between a $4\mu m \times 4\mu m$ graphite flake with a two-dimensional single crystallinured surface (Fig. 2a) and the n-Si surface with an atomical smoothness (Fig. 2b), the whole experiment method is shown in Method section and Supplementary Section 5. The measured friction forces under various normal forces are plotted in Fig. 2c, showing an extralow friction coefficient of 0.0039~0.0045. Furthermore, we cunducted an 6,000 circles sliding experiment for $4\mu m \times 4\mu m$ graphite flake on n-Si, the measured friction forces of the whole process are shown in Fig. 2d. It can be seen that the friction forces has basically stabilized in the range of $0.2~0.3\mu N$ after the decline process [31] in the first around 600 cycles, which indicate that the SSL state between graphite flake and n-Si can stably exist for large sliding cycles. After the sliding experiment, we characterized the slided n-Si and graphite flake interface, where the morphology characterization result is shown in Fig. 2e, we can calculate the roughness before and after 6,000 cycles sliding are ~112 pm and ~180 pm, respectively, which indicate there is no visible wear of slided n-Si surface. And the Raman characterization results at the different positions (points 1-9 shown in the illustration of Fig. 2f) of slided graphite flake interface are shown in Fig. 2f, there is no observable D peak (1350 cm$^{-1}$), which indicate that there is no visible damage of graphite flake interface after 6,000 cycles sliding. Therefore, the above result confirms the wearless of SSL contact state between graphite flake and n-Si.

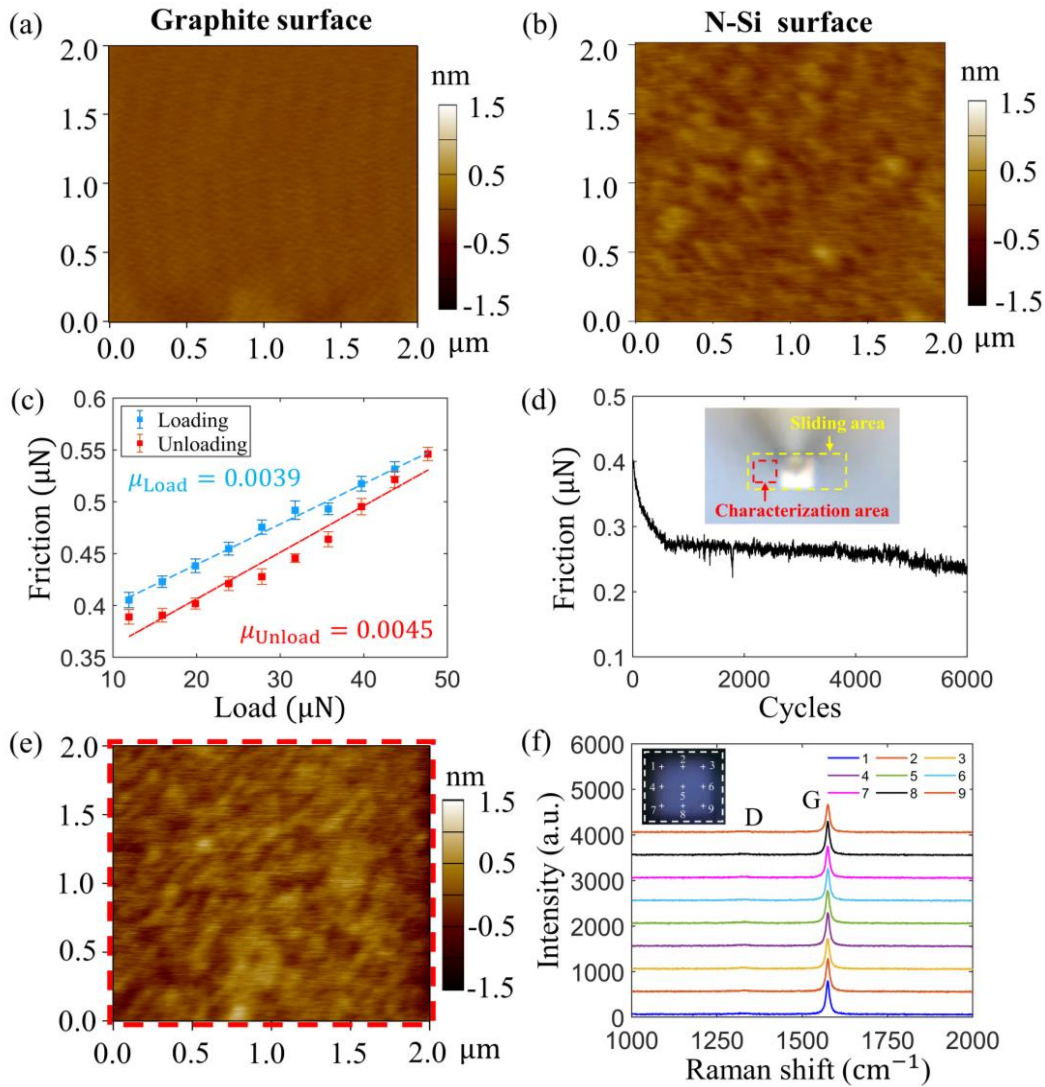

**Fig. 2 | The tribological tests between graphite flake and n-Si: a**, The morphological characterization of the two-dimensional single crystallinured surface of graphite flake. **b**, The morphological characterization of the n-Si surface. **c**, The friction force under different normal forces with displacement amplitude 2μm and speed 4μm/s, where the blue points and red points correspond to the loading process and the unloading process, respectively, each point was tested 40 times, which the friction coefficients are 0.0039 and 0.045, respectively. **d**, The measured friction force during the continuous 6000 cycles sliding process with displacement amplitude 4μm and

speed 8μm/s under a normal force of $N = 23.8\ \mu N$. The illustration is a optical image of the experimental setup, and the yellow dashed frame is the sliding region. **e**, The morphological characterization of n-Si interface after 6000 sliding cycles, the red dashed frame in **d** is the characterization region. **f**, The Ramman characterization of the graphite flake interface after 6000 sliding cycles, and the points 1-9 in the illustration represent the test posituion.

Now we show that the friction in our SLSG is too small to excite the electrons. According to the friction measurement of graphite/n-Si SLSG in Fig. 1d, the maximum friction force during the sliding process is about $f_{max}^{(1)} = 0.55\mu N$. Recently, it was proved that the friction in a superlubric contact is mainly occurring along the edges through the dangling bonds [36]. By approximating the friction force to be evenly distributed along the edge, we estimate the friction force of each dangling bond, $f_0^{(1)}$, to be is: $f_0^{(1)} \approx f_{max}^{(1)} a/4L = 8.5$ pN, where $a = 0.246$nm is the lattice constant of two-dimensional graphite surface. Futher, we estimate the upper limit of friction energy $\Delta E_f^{(1)}$ generated by the interaction of each dangling bond with silicon atoms as $\Delta E_f^{(1)} \approx f_0^{(1)} b = 0.0287$ eV, where $b = 0.543$ nm is the lattice constant of silicon. By noting $\Delta E_f^{(1)} \ll \Delta E_g = 1.12$ eV, where $\Delta E_g$ is the band gap of silicon, we conclude that the mechanism of graphite/n-Si system SLSG is very impossible to be the friction excitation.

In contrast, according to the friction measurement of AFM tip/n-Si TSG in Fig. 1f, the minimum friction force during the sliding process is about $f_{min}^{(1)} = 1\ \mu N$. According to the Hertz's contact theory [37,38], we can get the maximum contact normal stress to be 1.5 times the average contact normal stress. By assuming that the friction shear stress is proportional to contact normal stress at the contact area $A$ of the tip, we can obtain the estimate of the maximum friction shear stress as $P_f = 1.5 \times f_{min}^{(1)}/A \approx 4.37$ GPa, the Supplementary Section 4 gives futher details. Accordingly, the lower limit friction energy $\Delta E_f^{(2)}$ is $\Delta E_f^{(2)} = P_f c^2 b \approx 2.46$ eV, where $c \approx 0.408$ nm is the lattice constant of Au. The result of $\Delta E_f^{(2)} > \Delta E_g$ confirms that the friction excitation should be the main mechanism of AFM tip/n-Si system TSG.

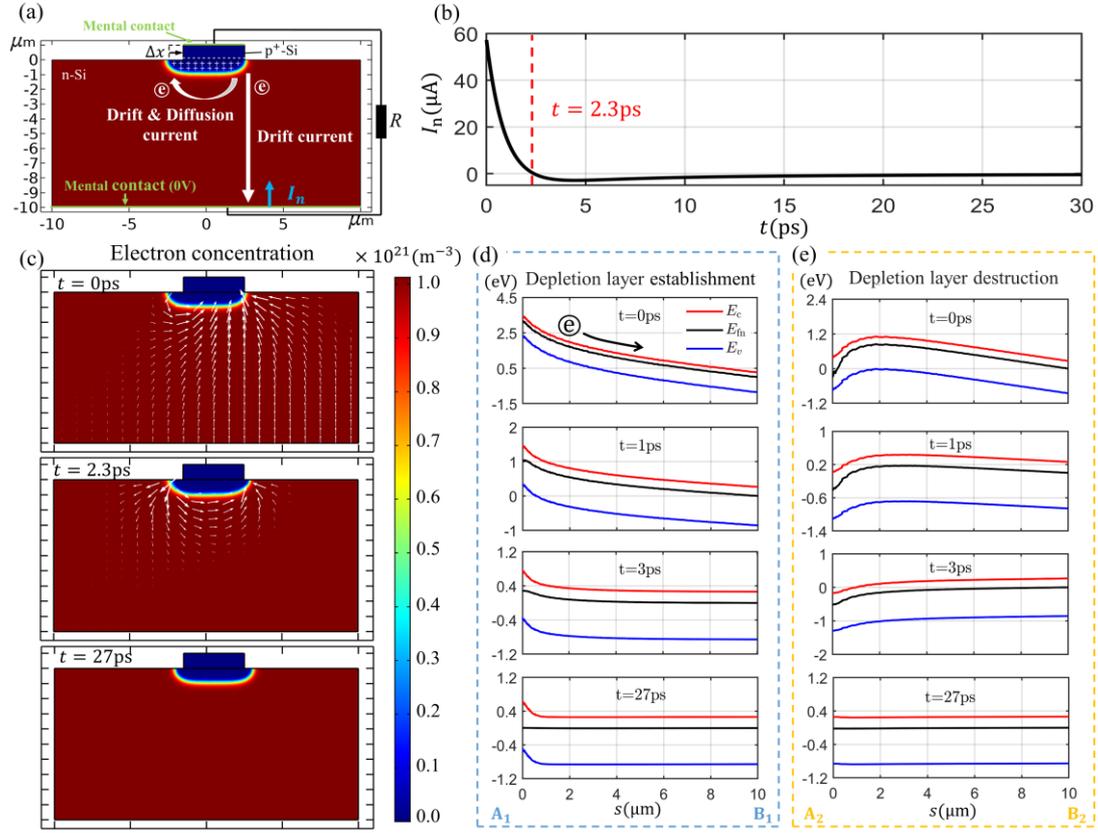

**Fig. 3 | The quasi-static semiconductor finite element simulation to explain the physical process of the EDDL mechanism when $\Delta x = 0.5\mu m$: a**, The model structure diagram and physical process. **b**, The relationship between the output electron current along the bottom surface of N-Si. **c**, The relationship between the electron concentration distribution over time, where the white arrow represents the direction of the electron current, that is, the reverse direction of the electron motion. **d**, and **e**, are the distribution of the conduction band energy level (red), valence band energy level (blue) and quasi-electron Fermi level (gray) at the $A_1B_1$ cut line ($x = 2.6$ μm, corresponding to the position of the depletion layer establishment) and $A_2B_2$ cut line ($x = -2.1$ μm, corresponding to the position of the depletion layer destruction) in **a**, respectively.

In addition to the friction excitation mechanism, there is another conjectured mechanism of Schottky generators called depletion layer establishment and destruct (DLED) [14,15]. When a metal contacted with an n-Silicon (n-Si), electrons will transfer from the n-Si to the metal [39]. Consequently, a positively charged region is formed in the n-Si which is the depletion layer, and the built-in electric field is generated accordingly. In the DLED mechanism it is assumed that sliding the metal will cause the head depletion layer to establish and the tail depletion layer to destruct, and thereby form a built-in electric field separation, which would cause the diffusing carriers emit to form DC current [14,15]. Nevertheless, there were neither experimental, nor theoretical proof of the DLED mechanism until the present work. To explain the experimental observation of our SLSG, we have excluded the friction excitation mechanism in the aforementioned part of this Letter. Hereinafter we show that the DLED is more likely the mechanism of our SLSG.

Since it is still lack of analyzing the electronic dynamic transportation behavior of the continuous sliding contact in SLSG, we perform a quasi-static finite element simulation. As illustrate in Fig. 3a, the model consists of a slider (length 4μm and height 1μm) at the top and a much larger stator (length 20μm and height 10μm) at the bottom. The stator is made of n-Si with a doping concentration of $N_\mathrm{D} = 10^{15} \mathrm{cm}^{-3}$. The slider is made of a heavily doped p$^+$-Si as an equivalent metal (The depletion layer penetrates primarily into the n-Si, and the width of depletion in the p$^+$-Si can be neglected, which is similar to metal/semiconductor contact, the Supplymentary Fig. 10 gives futher details). The top surface of the slider and the bottom surface of the stator are conductively connected through an external circuit series with a resistance $R = 10\mathrm{k\Omega}$ (the simulation results under different resistances are shown in Supplementary Fig. S13), and the potential of the bottom surface of the stator is set to be 0. Firstly, we simulate the electron and hole distribution when the Schottky diode is formed by the slider and stator (The results are shown in Supplementary Fig. S10). After equilibrium, we move the upper slider with a displacement of $\Delta x = 0.5\mathrm{\mu m}$ with a contrived constraint that the electron distributions of both the slider and the stator would not change, as shown in Fig. 3a. When removing the constraint, the unbalanced electric field will drive the electrons to flow along the external circuit and inside the slider and stator, and finally reach a new equilibrium state. Fig. 3b shows the output electron current $I_\mathrm{n}$ along the bottom surface of stator with time. It can be seen that there is a large current in the short period of time ($t < 2.3$ ps) when the simulation starts with the direction of the current as shown by the blue arrow in Fig. 3a, which is opposite to the direction of electron movement and consistent with the direction of current measured in Fig. 1. The current will change the direction at $t = 2.3$ ps and gradually decay to zero. In order to further explain the cause of this current, we draw the electron concentration distribution at different time in Fig. 3c (the electric potential distribution is shown in Supplementary Fig. S12), where the white arrow represents the direction of the electron current, that is, the reverse direction of the electron motion. The conduction band energy level (red), valence band energy level (blue) and electron quasi-fermi level (gray) at $A_1B_1$ and $A_2B_2$ in Fig. 3c are shown in Figs. 3d and e, corresponding to the position of the establishment ($x = 2.6$ μm) and destruction ($x = -2.1$ μm) of the depletion layer, respectively (the electric field in the y direction of $A_1B_1$ and $A_2B_2$ are shown in Supplementary Figs. S12b and c). When $t = 0$ ps, the separation of build-in electric field will bend the energy band of $A_1B_1$ and $A_2B_2$ at the bottom surface of stator, which cause the electrons to drift and move out along the bottom surface of stator. Therefore, $I_n$ at $t = 0$ ps is mainly contributed by the drift motion of the electron due to the unbalanced electric field, and the degree of the energy band bending decreases when $t = 1$ ps shown in Figs. 3d and e, which explains the attenuation of $I_n$ when $0$ ps $< t < 2.3$ ps in Fig. 3b. When $t > 2.3$ ps, that is the energy band in $t = 3$ ps shown in Figs. 3d and e, the energy band bending near the bottom surface of the stator disappears, and only appears near the depletion layer, that is, the electrons are mainly transferred inside to reach equilibrium rather than the external circuit. At last, when $t = 27$ ps, as shown in Figs. 3d and e, the whole region reaches a new equilibrium state, which $I_n$ decays to zero, and the energy band of the $A_1B_1$ and $A_2B_2$ cut lines converges to a new static equilibrium distribution (the quasi fermi levels are equal).

In sum, the above simulation gives a physical image to verify the feasibility and rationality of the EDDL mechanism. This physical image indicates that the output current is mainly contributed by the electron drift motion caused by the non-equilibrium electric field during the movement of slider. The relaxation time of the simulation process is much shorter than the characteristic time of the movement, which means the main contribution of the output current is the drift current corresponding to the initial time $(I_n(t=0))$ in Fig. 3b when we give a continuous movement to slider.

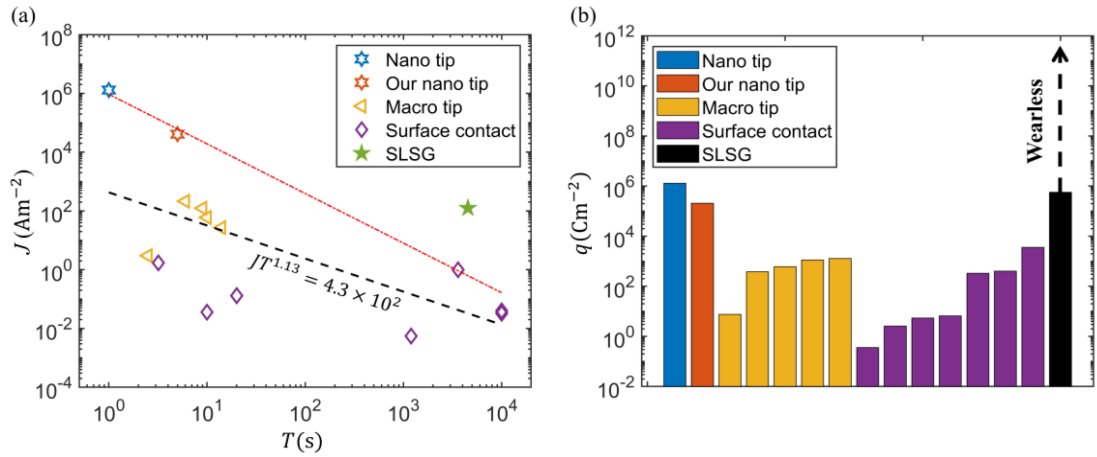

**Fig. 4 | The comparison of SLSG and all reported TSGs** [3,14,16,19-21,33,40-42]: **a**, The relationship between current density $J$ and lifetime $T$, where different shapes and colors of dots represent different contact mode, and the black and red dotted line indicate the power exponential fitting of all points and three upper right points of reported TSG, respectively. **b**, The transfer charge density $q$, which is the integral of the current density over the time in lifepan ($q = \int_0^T J dt$), of all reported TSG and SLSG.

Before conclude this paper, we make a comparison between SLSG and all other reported TSGs to illustrate the advantages of SLSG. First of all, we divided the all reported TSGs into nano tip [3,16], macro tip [3,16,33,40,41] and surface contact [14,19-21,42] TSGs according to its contact area (more details describtion are shown in Supplementary Section 7), and the statistic current density $J$ and working lifetime $T$ reported in literature are shown in Fig. 4a. It can be seen that there is a contradiction between the current density and life time of all reported TSGs. For example, the nano tip [3,16] TSG with highest current densities tend to have lowest lifetime, and our experiments in Figs. 1e and 1f are also confirm this feature (red point ploted in Fig. 4a). Futher, we fit all the points and three upper right points of reported TSGs to get the black and red line, respectively, and the point obtained by the SLSG experiment is at the upper right of the two lines, which means that it has relied on this contradiction and achieved the both high current density and lifetime. Secondly, we integrate the current density over time in lifepan to get the transfer charge density $q = \int_0^T J dt$ of the all reported TSGs and SLSG, shown in Fig. 4b, which indicate the working capacity. It can be seen that SLSG has larger transfer charge density compare to mostly reported TSGs, and we believe that the achievable transfer charge density of SLSG should be larger or substantively unlimited because no current decay and wear were observed during the entire experiment.

To conclusion, we demonstrate the first prototype of a superlubric nanogenerator, namely a superlubric Schottky generator (SLSG) in microscale, that was made of a microscale graphite flake and an n-type silicon (n-Si). This SLSG can stably generate a DC electrical current for at least 5,000 cycles with a high density of $\sim 119 \text{ Am}^{-2}$, which is three orders of magnitude higher than those of any TENGs (the highest reported current density is $\sim 0.1 \text{Am}^{-2}$ [27]) and PENGs (the highest current density is $\sim 0.01 \text{Am}^{-2}$ [28]). Further, the fact that no observable wear and and current decay in the SLSG were found may imply substantially unlimited life of the SLSG. Our SLSG is thus the first nanogenerators that can have a high and staby efficiency and ultralong or unlimited life. In comparison, although Schottky generators based on friction excitation can generate momentary higher current density ($10^4 \sim 10^7 \text{ Am}^{-2}$), they paid the cost of high wears and consequently very short lives (a few to a dozens of circles). In addition, our work provides also a direct and experimental proof of the conjectured EDDL mechanism, and a theoretical and quantitative explanation of the EDDL process. These experimental and theoretical results may guide and accelerate the SLSG into real applications.

# Method

**Preparation of graphite-Au flake.** We firstly fabricate square graphite mesa arrays with Au film on highly ordered pyrolytic graphite (HOPG, ZYB grade (Brucker) [43]). The fabrication process is shown in Supplementary Fig. S1a, we firstly spin on a layer of photoresist on the fresh cleavage surface of the HOPG (i), and then remove the photoresist of the mesa array area by electron beam lithography (ii). Secondly, we obtain the Au array film with thickness of 100nm through electron beam evaporation (iii) and lift-off (iv) process. Lastly, we use metal as a mask to obtain graphite mesa with Au film by reactive ion etching (oxygen ions) process (v), where the etching depth is 2.5 μm. The characterizations of fabricated graphite mesa with Au film are shown in Fig. S1b. In order to form the SLSG structure shown in Fig. 1a, we need to transfer graphite-Au flake with single crystal superlubric interface to the n-Si surface, and the specific process is shown in Supplementary Fig. S2 and Section 2. We use a tungsten microtip controlled by a micromanipulator (Kleindiek MM3A) as validated by optical microscopy (HiRox KH-3000) to apply a shear stress until they split along their vertical direction, We determine whether the sheared graphite-Au flake undergoes self-recovery motion (SRM) [29] to determine whether it has a single crystal superlubric interface [30].

**Preparation of n-Si.** For the fabrication of n-Si, we use electron beam evaporation to deposit a layer of 100nm Al on one side of a 4-inch, 200um thick, double-polished silicon wafer with a <100> crystal plane, then use the wafer scriber to cut into small pieces of 1cm × 1cm size, soak the each piece in BOE (Buffered oxide etch) solution for 15 minutes to remove the oxide layer on the surface of n-Si, and finally clean with acetone, alcohol, and deionized water, encapsulate with vacuum.

**SLSG formation.** We use the microtip to drag the dangling graphite flake controlled by the micromanipulator (Kleindiek MM3A), and place it slowly by micromanipulator on the atomically smooth fabricated n-Si surface, at this time, since the adsorption force of the graphite flake and n-Si is larger than that of the microtip and graphite flake, the graphite island will remain on the n-Si surface, as shown in Supplementary Fig. S2f, which form the SLSG structure shown in Fig. 1a. In order to verify that the contact formed by the graphite-Au flake and n-Si is Schottky contact, we performed a static current-voltage (I-V) characteristic measurement on the transferred SLSG structure in Supplementary Fig. S2f through NT-MDT AFM system. The detailed experimental and fitting results are shown in Supplementary Fig. S3.

**Friction and current measurements.** The friction and current measurements of the graphite/n-Si heterostructures were performed under an ambient atmosphere (temperature of 25 °C and relative humidity of 29 %). The experimental set-up included a commercial NTEGRA upright AFM (NT-MDT), a 100 μm XYZ piezoelectric displacement platform, a high numerical aperture objective lens (× 100 (Mitutoyu)) and visualized conductive AFM tip (ACCESS-NC-GG(Appnano)). Fig. 1a shows the schematic of the experimental set-up. We can accurately press the AFM tip on the Au cap of the graphite flake through the optical microscope. The AFM tip was calibrated in situ by the Sader method [44,45] for the normal direction force and the diamagnetic levitation spring system [46] for the lateral direction force, the specific results of the calibration process are shown in the Supplementary Fig. S4. The bottom Al film of the n-Si is grounded through the iron stage, and the conductive AFM tip is also grounded by connecting a precision ammeter, which can accurately measure the current through AFM tip in the sliding process, and the noise current measurement of NT-MDT AFM system is shown in Supplementary Fig. S5a, which is basically maintained at the order of 1pA.

**The contact area calculation of AFM tip/n-Si TSG.** For AFM tip/n-Si TSG, the DMT model can best approximate the contact between the AFM tip and the hard poorly adhesive material. According to the DMT model [32], the contact area $A$ is given by:

$$A = \pi \left( \frac{R}{K} (F_N + 2\pi R \gamma) \right)^{\frac{2}{3}}, \qquad (1)$$

where $R$ is AFM tip radius, $F_N$ is the normal force apply to the AFM tip, $\gamma$ is the energy of adhesion, the term $2\pi R\gamma$ can be considered as an additional load, which is determined by the "pull-off" force in the force curve, and $K = \frac{3}{4}\left(\frac{1-\nu_s^2}{E_s} + \frac{1-\nu_t^2}{E_t}\right)$ is the reduced Young's modulus, where $E_t$ and $E_s$ are Young's moduli, and $\nu_t$ and $\nu_s$ are the Poisson ratios of the tip and the sample, respectively, and the calculation details as shown in Supplementary Section 4. Since the "pull-off" force is much smaller than the normal force apply to AFM tip ($2\pi R\gamma \ll F_N$), we can use Hertz contact model [37,38] replace the DMT contact model [32] to consider the normal pressure distribution, as details in Supplementary Section 4, the maximum friction shear stress of the contact region is $P_\text{f} = \frac{3}{2}\frac{f}{A}$, and $f$ is the measured friction force.

**Surface characterization method of tribological experiment.** Here, we will show the main process and method of the experiment in Fig. 2. The first step is to use the Asylum Research Cypher S AFM in tapping mode to characterize the topography of the interface (Fig. 2a) of selected graphite flake with SRM [29] property through flip 180 degree of microtip after the microtip adsorbs the graphite flake (Supplementary Fig. S2e) before placing it on the n-Si surface (Supplementary Fig. S2f), and characterize the topography of n-Si surface (Fig. 2b) at the same time. For the second step, we use the lateral force measurement system of AFM (NT-MDT) to measure the coefficient of friction (Fig. 2c), and then measure the friction force of 6,000 cycles sliding process (Fig. 2d). For the third step, we firstly use Asylum Research Cypher S AFM in tapping mode to perform a larger range morphological characterization and find the position of the graphite-Au flake as shown in Supplementary Fig. S8, that is, and further characterize the small sliding region of n-Si through the positioning function of AFM, to judge whether there is any observable damage (Fig. 2e). And we use the method shown in Supplementary Fig. S9 to lift the graphite-Au flake by overcoming the van der Waals adsorption force between graphite-Au flake and n-Si interfaces, and flip 180 degree of microtip to perform the Raman characterization on the flipped slided graphite-Au flake interface (Fig. 2f), to judge whether there is any observable damage from whether there is a D peak (1350 $\text{cm}^{-1}$).

**The setup of quasi-static simulation of DLED mechanism.** The geometric parameters and boundary conditions of the model are set as shown in Supplementary Fig. S10a, the model consists of a slider (length 4μm and height 1μm) at the top and a much larger stator (length 20μm and height 10μm) at the bottom. The stator is made of n-Si with a doping concentration of $N_\text{D} = 10^{15} \text{cm}^{-3}$. The slider is made of a heavily doped p⁺-Si with a doping concentration of $N_A = 2 \times 10^{19}$ cm⁻³ as an equivalent metal. is set to a continuous heterojunction condition

$$\begin{aligned} E_{fn}^{(1)} &= E_{fn}^{(2)}, \\ E_{fp}^{(1)} &= E_{fp}^{(2)}, \\ \vec{D_1} &= \vec{D_2}, \end{aligned} \quad (2)$$

where $E_{fn}^{(i)}$ and $E_{fp}^{(i)}$ ($i = 1,2$ represents the slider and stator) are the fermi levels of electrons and holes respectively. The top surface of the slider and the bottom surface of slider are set to metal contact boundary, which are conductively connected through an external circuit series with a resistance $R$, the potential of the bottom surface of stator is set to be 0, and the potential of the top surface of the slider $V_1$ is determined according to the continuous conditions of the external current $I_R$ flowing through the resistance

$$\begin{aligned} V_1 &= I_R R, \\ I_R &= \iint I_n dS, \end{aligned} \quad (3)$$

where $I_n$ is the normal current on the bottom surface of the stator. The potential distribution $V$ in the semiconductor satisfies the Poisson equation

$$\nabla \cdot (\varepsilon \nabla V) = q(n - p - N_D + N_A), \quad (4)$$

where $\varepsilon$ is the permittivity, $n$ and $p$ are the electron and hole concentration respectively. The relationship between carrier concentration and energy band is given by statistical theory

$$\begin{aligned} n &= N_c \exp\left(-\frac{E_c - E_{fn}}{kT}\right), \\ p &= N_v \exp\left(\frac{E_v - E_{fp}}{k_B T}\right), \\ E_c &= -\chi_{n-Si} - qV, \\ E_v &= -\chi_{n-Si} - E_g - qV, \end{aligned} \quad (5)$$

where $N_c = 2\left(\frac{2\pi m_e^* k_B T}{\hbar^2}\right)^{\frac{3}{2}}$ and $N_v = 2\left(\frac{2\pi m_h^* k_B T}{\hbar^2}\right)^{\frac{3}{2}}$ are the thermally excited state density of electron and hole, $E_c$ and $E_v$ are the conduction band and valence band energy levels, $E_g = 1.12$ eV is the band gap of silicon. By solving equations Eq. (4) and Eq. (5), we can get the potential distribution and carrier concentration distribution in the semiconductor, and further, we can calculate the electron and hole currents ($\vec{J_n}$ and $\vec{J_p}$) by the drift diffusion model

$$\begin{aligned}\vec{J_n} &= -qn\mu_n\nabla V + qD_n\nabla n, \\ \vec{J_p} &= -qn\mu_p\nabla V + qD_p\nabla p, \\ \nabla \cdot \vec{J_n} &= q\frac{\partial n}{\partial t}, \\ \nabla \cdot \vec{J_p} &= -q\frac{\partial p}{\partial t},\end{aligned} \qquad (6)$$

where $\mu_n$ and $D_n$ are the electron mobility and diffusion coefficient respectively, the $\mu_p$ and $D_p$ are the hole mobility and diffusion coefficient respectively.

**The simulation process.** The whole quasi-static simulation process will be divided into three steps: Step 1: we simulate the electron and hole distribution when the Schottky diode is formed to reach static equilibrium ($\Delta x = 0$) by the slider and stator. Step 2: After equilibrium, we move the upper slider with a displacement of $\Delta x = 0.5\mu m$ with a contrived constraint that the electron distributions of both the slider and the stator would not change, as shown in Fig. 3a and Supplementary Fig. S10a, and the carrier distribution has reached a non-equilibrium state. Step 3: We remove the constraint and start the transient simulation with the state of the step 2 as the initial condition, to simulate the transport process of carriers, electric potential distribution and related parameters (energy band distribution of cutline $A_1B_1$ and $A_2B_2$) over time.